# Object Detection and Geometric Profiling through Dirty Water Media Using Asymmetry Properties of Backscattered Signals

Chensheng Wu*, Robert Lee, Christopher C. Davis

Department of Electrical and Computer Engineering, University of Maryland College Park,
MD USA 20742

## ABSTRACT

The scattering of light observed through the turbid underwater channel is often regarded as the leading challenge when designing underwater electro-optical imaging systems. There have been many approaches to address the effects of scattering such as using pulsed laser sources to reject scattered light temporally, or using intensity modulated waveforms and matched filters to remove the scattered light spectrally. In this paper, a new method is proposed which primarily uses the backscattering asymmetry property for object detection and geometric profiling. In our approach, two parallel and identical continuous wave (CW) laser beams with narrow beam widths (~2mm) are used as active illumination sources. The two beams also have controllable spacing and aiming angle, as well as initial phase difference for convenience of scanning and profiling a target. Through theory and experimental results, it will be shown that when an object leans or tilts towards one of the beam's central trajectory, the asymmetry in the backscattered signals can be used to indicate the location or slope of the target's surface, respectively. By varying the spacing or aiming angle of the two beams, a number of surface samples can be collected to reconstruct the object's shape geometrically. The resolution and range limit of our approach are also measured and reported in this work. In application, our proposed method provides an economic solution to perform imaging through turbid underwater environments. Additionally, the idea can be combined with the pulsed or modulated laser signals for enhanced imaging results.

## 1. INTRODUCTION

There are many potential techniques to perform enhanced vision or object detection in a dirty/turbid water environment [1-3]. Many of them could require a specific Lidar system based on the time of flight (TOF) concept [4-9], or a hybrid system with conventional acoustic devices [10-12]. For economic reasons, the cost accompanied with these well-known ideas prevents their wide spread use on underwater vehicles [13]. The studying of the scattering medium itself, however, has aroused great attention in seeking low cost solutions for imaging problems in a turbid water environment [14, 15]. Conceptually, all approaches are based on finding ways to differentiate the scattered signal from the true target return using various filtering methods [16, 17]. Following the same philosophy, it is proposed that an unmodulated CW laser is capable of providing a much enhanced vision through the dirty water with little given or empirical knowledge of the scattering medium.

In our method, a thin CW laser with beam width < 3mm is split into two identical beams through a 50-50 non-polarizing beam splitter and co-propagated through the underwater medium in parallel. The spacing between the two laser beams is adjustable and typically set within the radius where the cross section of the forward scattering pattern becomes fully speckled. Empirically, the spacing can be set around 2~20 mm to perform general imaging tasks without optimization. The spacing also determines the smallest sub-area of the target that is detectable. When the sub-area of the target is facing towards the injected beams at a normal incidence angle, the back scattered light from the two beams will have identical patterns. If the sub-area tilts towards one of the twin beams, asymmetry of the returned pattern can be observed as the propagation depth, transverse coherence length, and spread angle of the beams will be different. The asymmetry pattern can be translated to the angle of surface tilt under a general assumption that the sub-area of the target is regionally flat and uniform in reflectivity. Each gradient sample (sub-area tilt) requires at least one camera image. With the collection of different surface gradients retrieved through a scanning pattern of the twin beams, one can reconstruct the fundamental surface profile of the target.

Compared with the normal visibility limit achieved by using bright light from a LED or lamp for active illumination in the dirty water medium, the twin beam method can extend the range of object recognition by a factor between 3~5.

Given that such a system only requires one CW laser, one camera, one beam splitter and several steering mirrors that fit within a self-contained enclosure, it features a cost-effective solution to many underwater imaging systems for detection or navigation purposes.

## 2. WORKING PRINCIPLE

### 2.1 Assumptions and modeling simplifications

For simplicity, it is assumed that the highly scattering medium has a uniform size parameter $\chi=(2\pi n_{med}\alpha)/\lambda_0$, which is a commonly used quantity in most scattering analysis [18]. In reality, the refractive index of the scattering particle may vary and be a mixture of multiple types of particles. Correspondingly, the statistical average number $n_{med}$ or the dominant particle's quantity $n_{med}*$ can be used in our simplified theory to provide simple evaluation of our proposed method.

It is necessary to assume that the proposed method should be operated where single scattering is the dominant cause of near axis attenuation [19]. This is typically satisfied in most natural situations, where in rare cases with dominant multiple scatterings our method will be invalid as a significant amount of side scattered light with large angles may be routed back to the forward scattering direction to degrade the quality of detection. Under this assumption, a closed form of point spread function (PSF) in small-angle approximation of radiative transport equation can be written, and the beam spread can be expressed as the result of accumulated small-scattering events as [20, 21]:

$$\langle r^2 \rangle = \frac{4\pi}{3\chi^2} sz^3 \ . \tag{1}$$

In Eq. (1), $s$ is the volumetric scattering coefficient and $z$ represents the propagation length. Although studies in the small-angle approximation for volume-scattering function (VSF) and point-spread function (PSF) are not perfectly consistent with data, it still features a simple and good description of how image resolution would be affected by a scattering medium [22]. Obvious, Eq. (1) denotes that the spreading radius of a beam is primarily affected by particle size and propagation distance as:

$$R := \sqrt{\langle r^2 \rangle} \sim z^{1.5}/\alpha \ . \tag{2}$$

As the twin beams propagate through the turbid underwater medium and reach a sub-area on the target, a structured illumination pattern will be formed on this area. The amplitude contribution from each beam is assumed to be the sum of the spread beam and the unscattered beam that propagates near the axis. As the propagation distance typically stays well within a 100m range, the diffraction angle of the beam can be ignored, and the unscattered beam portion resembles the original beam with reduced amplitude. The irradiance function can be typically written as (by adopting Arnush's deduction in [20]):

$$h(r,z) \sim g e^{-sz} \left(\pi R^2\right) \exp\left(-\frac{r^2}{R^2}\right) I_{r>r_0} + \exp\left(-\frac{r^2}{w_0^2}\right) I_{r \le r_0} \ . \tag{3}$$

In Eq. (3), $I$ is used as an indicator function and $w_0$ is the beam width (~2mm). The irradiance is expressed in an unscaled form with a coefficient $g$ indicating the relative strength between the scattered and unscattered beam parts. Note that Eq. (3) is also a simplification for beam profile in the forward direction, which should only be used for conceptual explanations.

Additional simplification needs to be applied to the returned scattering pattern. As the camera is pointing towards the forward propagation direction and the propagation distance z is much larger than the camera diameter, reciprocity principles can be used [23] and the small angle approximation (SAA) can be implemented for the returned beam pattern. Unless the target has a mirror-like surface, the returned scattering pattern is typically the form of Eq. (3) in convolution with its first part under a Lambertian surface assumption. The reflected pattern can be written as:

$$h(r,0) = e^{-sz}\left(\pi R^2\right)\exp(-\frac{r^2}{R^2}) * \left[h(r_1,z_1) + h(r_2,z_2)\right] \tag{4}$$

In Eq. (4), the forward irradiance of the twin beams are expressed with local transverse coordinates $r_1$ and $r_2$ that are referenced to their beam centers respectively.

## 2.2 Image acquisition process

When an image is taken by the camera with a short exposure time, the light field that enters the camera contains the forward scattering part from the target plus the backward scattering part at each depth before the twin beams reach the target in forward propagation. The first part is simply Eq. (4), and the second part is proportional to the integral of the convolution results of Eq. (3) and its first part. In general, this can be expressed as:

$$I^+(r,0;z) = \gamma e^{-sz} \left(\pi R_z^2\right) \exp(-\frac{r^2}{R_z^2}) * \left[h(r_1, z_1) + h(r_2, z_2)\right]$$
$$+ \gamma \delta \int_0^z h(r_1, t) * e^{-st} \left(\pi R_t^2\right) \exp(-\frac{r_1^2}{R_t^2}) + h(r_2, t) * e^{-st} \left(\pi R_t^2\right) \exp(-\frac{r_2^2}{R_t^2}) dt. \quad (5)$$

In Eq. (5), the subscripts on $R$ are used to denote that it is a distance dependent parameter. The ratio of real field irradiance to pixel intensity on the camera is simply expressed by the coefficient constant $\gamma$. The relative back scattering coefficient is represented by $\delta$. As the focus of the approach is the asymmetry in the returned light pattern, the expression for the asymmetry pattern has been hypothetically created by tracking the difference between the twin beams as:

$$I^-(r,0;z) = \gamma e^{-sz} \left(\pi R_z^2\right) \exp(-\frac{r^2}{R_z^2}) * \left[h(r_1, z_1) - h(r_2, z_2)\right]$$
$$+ \gamma \delta \int_{z_2}^{z_1} h(r_1, t) * e^{-st} \left(\pi R_t^2\right) \exp(-\frac{r_1^2}{R_t^2}) - h(r_2, t) * e^{-st} \left(\pi R_t^2\right) \exp(-\frac{r_2^2}{R_t^2}) dt. \quad (6)$$

Intuitively, Eq. (6) is a mimic of Eq. (5) by flipping the "+" sign between the twin beams terms to a "-" sign. Because the twin beams share a major propagation path, the concentration, size and distribution of the scattering particles should be identical with each other. Eq. (6) can be expanded with the twin beams' differential part as:

$$I^-(r,0;z) = \gamma e^{-sz} \left(\pi R_z^2\right) \exp(-\frac{r^2}{R_z^2}) * \vec{d} \cdot \vec{\theta} \left[\frac{\partial h(r,z)}{\partial R} \cdot \frac{dR}{dz} + \frac{\partial h(r,z)}{\partial z}\right]$$
$$+ \gamma \delta \int_{z_2}^{z_1} h(r_1, t) * e^{-st} \left(\pi R_t^2\right) \exp(-\frac{r_1^2}{R_t^2}) - h(r_2, t) * e^{-st} \left(\pi R_t^2\right) \exp(-\frac{r_2^2}{R_t^2}) dt. \quad (7)$$

It is of great interest to find that if the sub-area surface of the target has a small tilt (represented by $\theta$), the second term of Eq. (7) won't contribute to the $I^-$ term while the first term will become dominant. In other words, because the effective optical thickness between the "landing depths" $z_1$ and $z_2$ of the twin beams are very small, there is a very trivial difference between the twin beams' back scattering pattern generated through their forward propagation process. By combining Eq. (7) and (3), the $I^-$ term is simplified to:

$$I^-(r,0;z) \approx \gamma e^{-sz} \left(\pi R_z^2\right) \exp(-\frac{r^2}{R_z^2}) * \frac{\vec{d} \cdot \vec{\theta}}{z} e^{-sz} \left(\pi R_z^2\right) \exp(-\frac{r^2}{R_z^2}) \left[3\frac{r^2}{R^2} + 3 - sz\right]. \quad (8)$$

In Eq. (8), the term $sz$ is commonly referred to as the optical thickness of the medium $\tau := sz$. When the spacing between the twin beams is aligned with the tilting direction, the whole asymmetry pattern will be proportional to the angular tilt. In other words, for any optical thickness, the angular tilt information will be carried by the entire asymmetry pattern. The asymmetry pattern versus the optical thickness of the media can be plotted as:

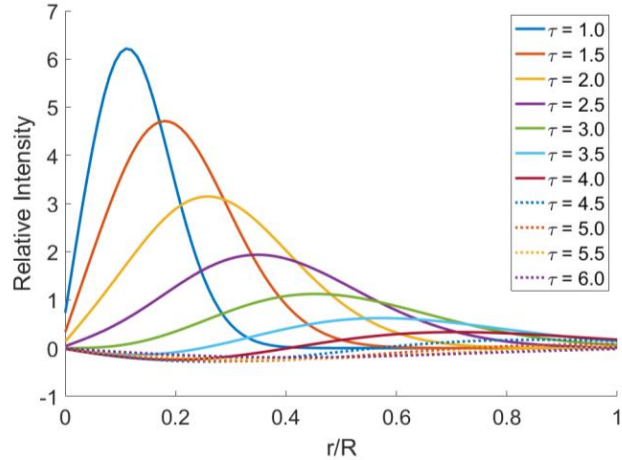

Figure 1. Asymmetry pattern between the two beams under a fixed optical thickness value

Figure 1 shows how the asymmetric intensity patterns vary at different radii. The best indication of asymmetry is found at the radius where the asymmetry pattern reaches a maximum. As the peak location shifts out monotonously with increased optical thickness, one can simply use the asymmetry pattern to estimate $\tau$ and obtain the local image section (typically a donut area) where the asymmetry is best performed to evaluate the surface tilt. It is also reflected by figure 1 that the proposed asymmetry method should work well within the range of $\tau=1\sim4$. For lower numbers of optical thickness ($\tau<1$), the peak location will be found within the beam width of the unscattered part, which prevents the method from obtaining a reliable result for the local surface tilt. For larger numbers of optical thickness ($\tau >4$, which implies longer distances), the curves flatten out with their magnitude dropping below a general speckled pattern. It is also worth mentioning that the SAA approximation is reasonably accurate for $\tau<5$ [24], which lines up very well with our working range. In other words, even if there is an integral method to enhance the asymmetry results, the approximations are no longer valid over $\tau>4.5$ as the fundamental theory itself will produce unexpected errors.

**2.3 Enhancing the asymmetry pattern for surface tilt extraction**

One can typically integrate over different radii where the asymmetry pattern beats the noise for enhanced results. This helps suppress the general speckled pattern in the asymmetry results, and allow us to see the real scale performance under various levels of optical thickness (primarily because the parameter $R$ in figure 1 is dependent on $\tau$). The noise is empirically estimated as $\sim 0.1$, and the enhancement factor (the summation of pixel values in the integral area per rad/cm tilt) through the integral method is shown as:

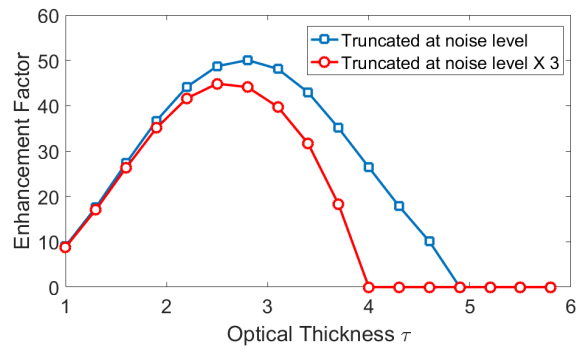

Figure 2: Enhancement factor of the asymmetry through area integral

Figure 2 shows that an enhancement factor >10 is generally achievable for the applicable range of the asymmetry method. Sometimes, it is safer to truncate at a higher level than the noise, which is shown through the red curve in figure 2. It is not surprising to see that at relatively close range (where $\tau\sim1$) and relatively far range (where $\tau>4.5$), the algorithm is not performing as well as in the medium range (where $\tau\sim2$ or $\tau\sim3$). This is because at the initial stage of beam propagation, there is not much beam spread so that asymmetry between the twin beams is locally confined. When the situation evolves into a relatively far range, the actual beam spread begins to degrade into speckle patterns and the beam has totally lost its transverse coherence. The optimal situation is in the medium ranges where the spread of the twin beams covers a much wider area (~$z^3$) in comparison to the near range and the forward scattered beams are still coherent.

### 2.4 Extraction of the asymmetry pattern: comparison between imaging the twin beams with odd & even frames and imaging the twin beams in one frame

The above theory is best fit if the beams are ON/OFF modulated with $\pi$ phase shift, and their images taken in odd and even frames, respectively. However, such method requires additional shutters for the ON/OFF control and synchronization mechanisms between the shutters and the camera system (the internal shutter of the camera). In practice, it is suggested to use co-imaging of the twin beams in one frame. It will be shown that by properly handling the overlay areas between the twin beams' back scattering patterns, the one frame method actually performs slightly better than the two frame method. In addition, as the proposed method will typically be combined with a scanning approach, it introduces additional cost to a practical system when the frame rate needs to doubled. Therefore, for engineering purposes, the one frame method is adopted for the rest of the work.

In the one frame method, one can typically find that the two peaks in the back scattered signal (in image $I^+$) indicate the center of the two transmitted beams. The image should be 180 degrees rotated using the center of the segment line that connects the two peaks. The modified asymmetry pattern $I^*$ can be extracted by the difference between the original image and the rotated image and pick any one of the peak centers for calculation. It is not surprising to find that $I^*$ performs slightly better than the original $I$ expressed in Eq. (8). This is the case because that for any arbitrary point x in the overlapping area and its counterpart point x* to be subtracted after the rotation, the original method extracts $I_1(x) - I_2(x^*)$ while the new method extracts $I_1(x)+I_2(x)-I_2(x^*)-I_1(x^*)$. The latter one is enhanced by the additional term of $I_2(x)-I_1(x^*)$ that typically has the same sign as $I_1(x)-I_2(x^*)$. This can be conceptually illustrated as:

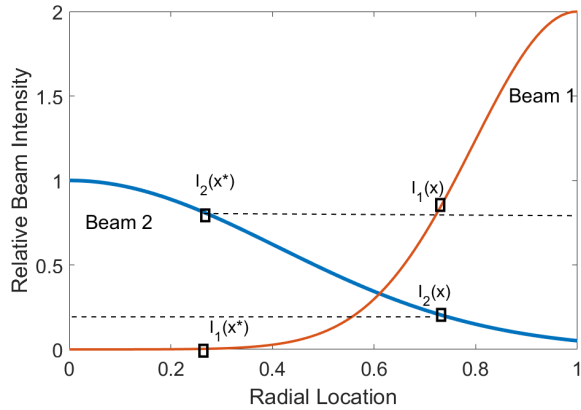

Figure 3: Illustration graph for the enhanced asymmetry results for I*

Based on the image processing algorithm, the slightly improved results can be shown as:

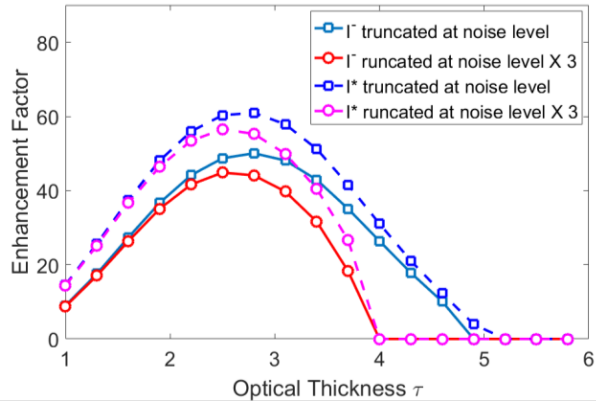

Figure 4: Demonstration of the slightly improved result through the one-frame method

In general, figure 4 shows that the method of using I* provides a 10% ~ 30% improvements in enhancement factor, and the improvement for seeing through certain optical thickness is trivial (or none). The latter observation can be explained as when the transverse structure of the beam breaks down at further depth in the scattering medium, the overlapping area performs equally "badly" as other non-overlapping areas.

## 3. EXPERIMENTAL VERIFICATIONS

Based on the theoretical analysis of the asymmetry property, the experimental platform is setup as:

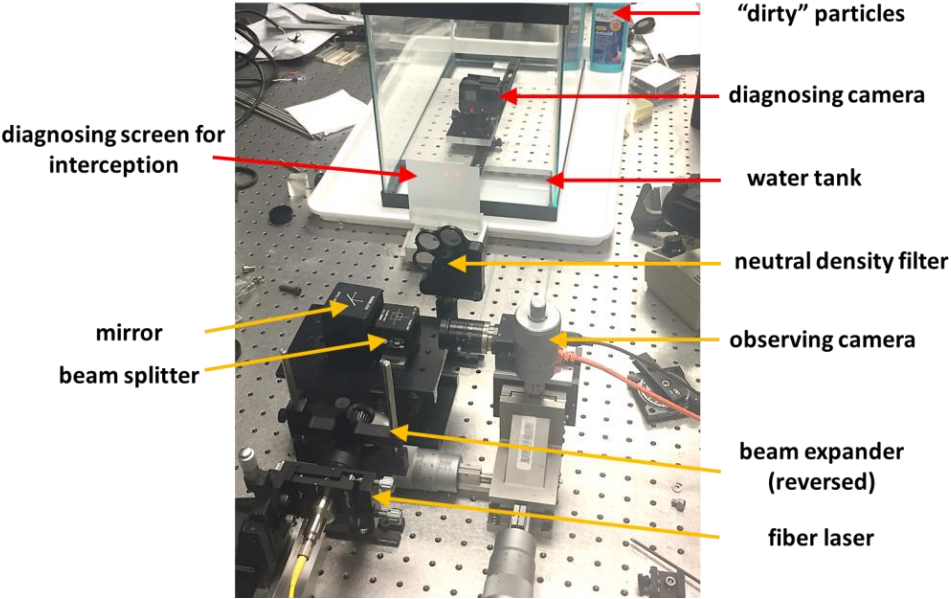

Figure 5: Experimental setup

In figure 5, the yellow arrows indicate the transmission and observation modules of the system, which all reside in the same sight. The red arrows indicate the system of the dirty water medium and parts for diagnosing purposes. A Gopro-Hero 5 camera was immersed in the water tank and used to evaluate the forward beam structure in correspondence to Eqs. (1-3). The water tank has a length of 0.5m, and the position of the camera can be indicated by the distance labels on an internal rail at the bottom of the tank.

### 3.1 Verification of SAA beam profile in forward propagation

The Go-pro camera was set with auto exposure time and narrow field of view (f~28mm) to image the beams as they propagate. Intuitively, this helps to validate that the experiments are conducted under the correct region of small angle approximation. A top-view camera was also used to image the beam spread and propagation path for comparison with the GoPro data. The sampled results can be shown as:

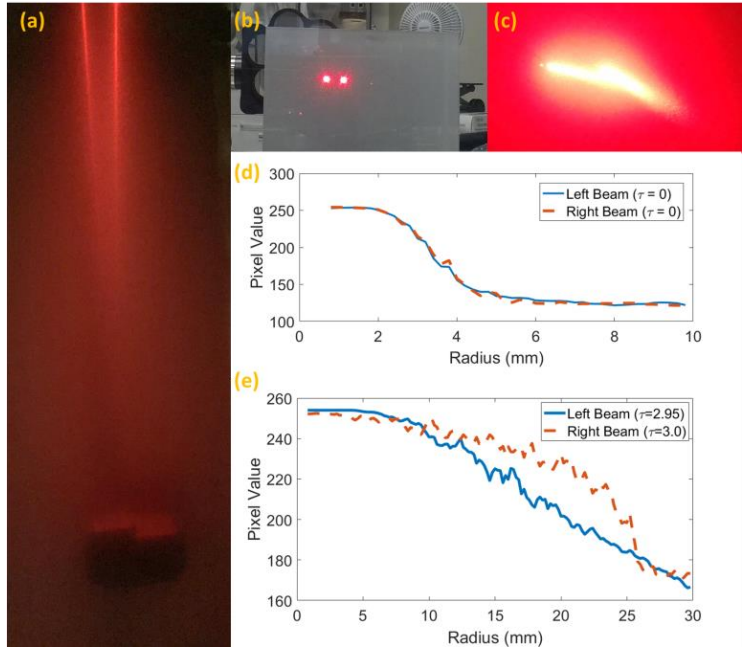

Figure 6: Sampled results in verifying the SAA through an imaging approach: (a) top view camera image; (b) initial image τ=0 intercepted through a diffuser screen; (c) GoPro image near τ=3; (d) averaged pixel value versus radius at τ=0 plane; (e) averaged pixel value versus radius near τ=3 plane;

In figure 6, it is clear to see that the small angle beam spreading preserves a Gaussian beam profile as it propagates. The top-view and GoPro suggested beam spreading radius in comparison with the theoretical predictions in Eq. (2) can be shown as:

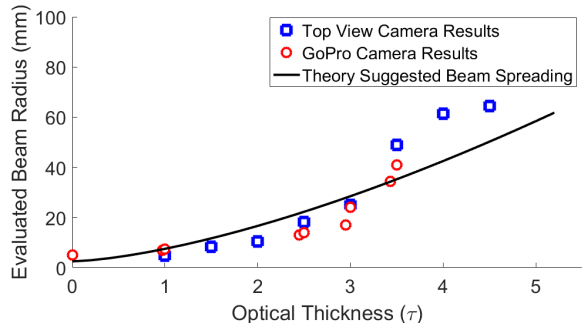

Figure 7: Evaluated beam spreading profiles through the imaging approaches

In figure 7, the camera evaluated beam spreading profiles suggest a steeper trend as opposed to theoretically predicted values at the same on-axis attenuation level (same optical thickness). This could be caused by a significant proportion of multiple scattering in the medium when the twin lasers penetrate through the turbid water. Overall, the SAA is essentially satisfied with the transverse beam intensity matching a Gaussian beam. The subtle mismatches in actual beam spreading width indicate that the proposed asymmetry method may expect a shorter truncation depth for effective object detection.

### 3.2 Verification of asymmetry properties

Without loss of generality, a spacing of 12 mm was used between the twin beams. A sandblasted aluminum structure featuring a triangular prism is used as a target object. The structure is mounted on a rotational stage with the vertical face aiming towards the twin beams. By rotating the mounting stage with a known angular tilt the asymmetry properties can be verified. For each optical thickness τ=1, 2, 3, the angular orientations of 0°, 5°, 10°, 20°, 40° and 80°, were tested.

The cases of 0°, 5°, and 10° tilting angles represent small surface tilts in general, while the other 20° 40° and 80° cases represent large surface tilts.

First, the asymmetry patterns' radial distribution was examined to compare the experimental results with the theoretical predictions shown in figure 1. The radial distribution aides in determining a rough estimate of the optical thickness $\tau_e$. Once the $\tau_e$ has been determined, the ring integral area can be determined with inner radius of $r_{min}$ and outer radius of $r_{max}$. The summation of pixel values in the ring area $r_{min} < r < r_{max}$ is proportional to the angular tilt and exposure time. In other words, the maximum asymmetry measured by the summation of the ring area can be combined with the camera exposure time to resolve the angular tilt of a hidden surface, expressed as:

$$\sum_{r_{min}<r<r_{max}} I_{asym}(i,j) = A_0 \kappa(\tau_e) \vec{d} \cdot \vec{\theta} \cdot T_{exp}. \tag{9}$$

In Eq. (9), $A_0$ is a constant coefficient for a fixed camera. $\kappa(\tau_e)$ is the enhancement factor from figure 4. In practice, $\kappa(\tau_e)$ should be experimentally determined to calibrate the theoretical predictions from SAA. $T_{exp}$ is the exposure time on the camera (typically expressed in units of micro-seconds).

When the tilting angle is 0°, there should be no asymmetry pattern according to Eq. (9). Therefore, the experiment should see a pattern that reflects the subtle noise in the system, which is plotted as:

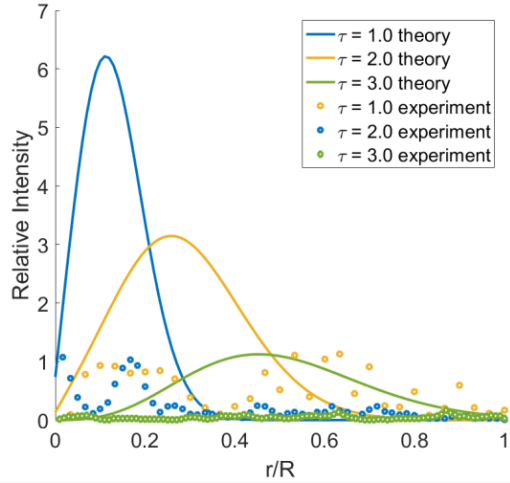

Figure 8: Subtle noise level reflected through the asymmetry pattern at a normal incidence angle

When a small tilting angle of 5° is given, the asymmetry pattern can be reflected as:

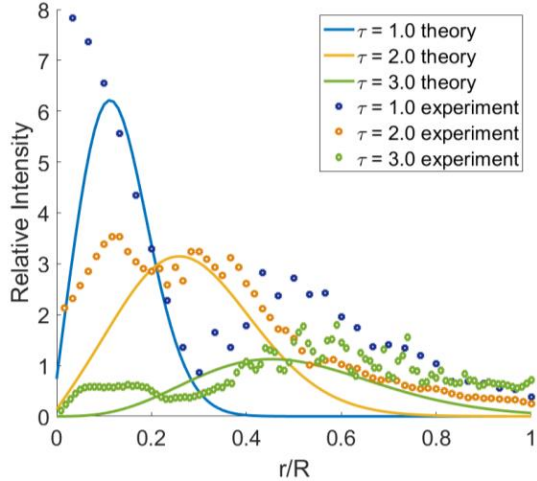

Figure 9: Experimental asymmetry pattern at 5 degrees surface tilt

Similarly, at surface tilt of 10°, the result can be shown as:

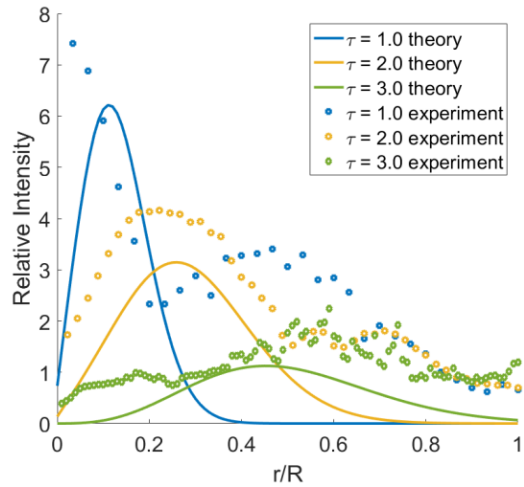

Figure 10: Experimental asymmetry pattern at 10 degrees surface tilt

Figure 9 and figure 10 shows a relatively good agreement between SAA theory and experimental data. The exception happens at small numbers of optical thickness ($\tau \sim 1$), where one additional peak shows up at further radius (around 0.5 R) indicating a larger asymmetry than theory prediction. For this reason, one can either use the $r_{max}$ from theory (as it agrees well with the first peak in experiment) or extend $r_{max}$ and use experimental enhancement factor.

At a larger surface tilt of 20°, the result is shown as (which is very similar to the previous cases):

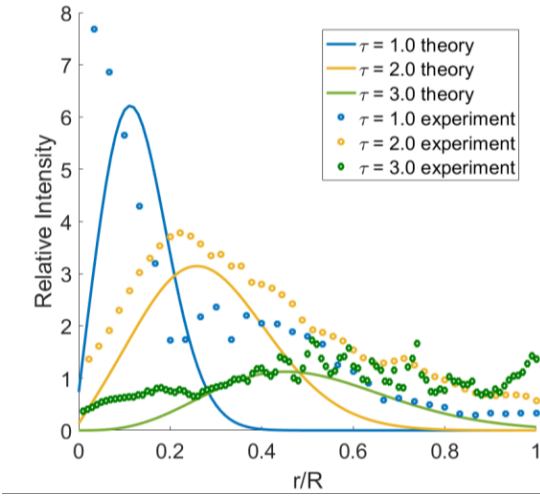

Figure 11: Experimental asymmetry pattern at 20 degrees surface tilt

Therefore, for a fairly large angle of 20°, the SAA perturbation result still agrees reasonably well with experimental data. And the exception happens when the tilting angle grows significantly larger than 30°. For example, the 40° data can be shown as:

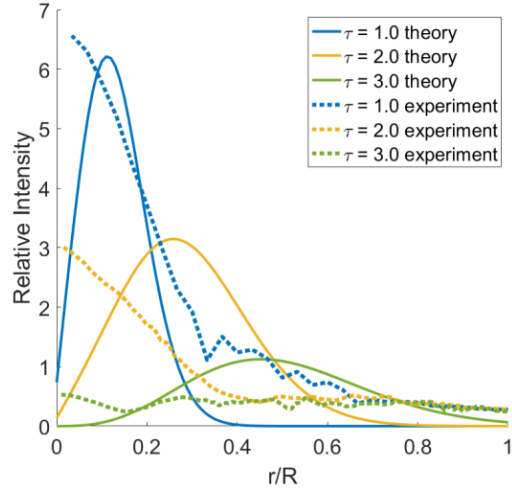

Figure 12: Experimental asymmetry pattern at 40 degrees surface tilt

Figure 12 suggests that at larger surface tilt, the asymmetry pattern is controlled by the fact that one of the twin beam's back scattering pattern strictly dominates over the other one's pattern at every radius. In this circumstance, one has to either reduce the spacing between the two beam, or use experimental data to determine $\kappa(\tau_e)$. The former will essentially meet a resolution limit (to be mentioned in conclusion) that one can no longer reduce the spacing for better results. The latter one requires a delicate look-up table.

For summary, we show the experimental statistics of the enhancement factor in the current twin beam format based on all integer angles from 1° to 30°. The result is plotted as:

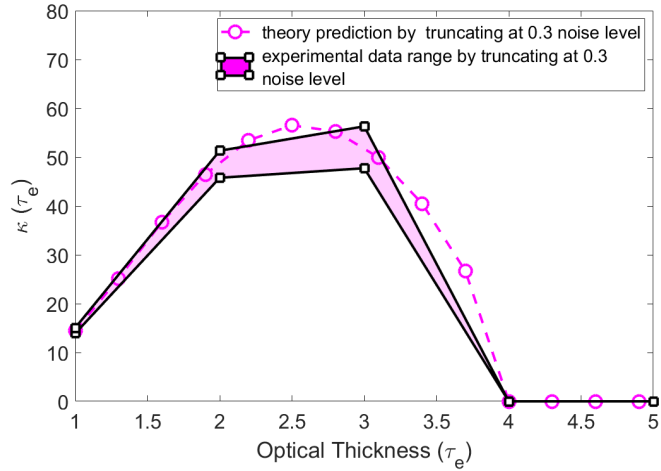

Figure 13: experimentally determined $\kappa(\tau_e)$

The truncation result for 0.3 noise level is used in correspondence to the experimentally determined asymmetry noise shown in figure 8. The statistics shown in figure 13 indicate that even without experimental calibration, the theoretical $\kappa(\tau_e)$ based on SAA will only cause an error around 20% for most cases. In other words, the theory can be directly applied if there is no strict accuracy requirement.

### 3.3 Imaging examples

To mimic a scanning pattern in reconstructing the surface of a target (the triangular prism), the beam is manually steered to project to a 5×6 equally spaced directions to extract 30 slope samples. Then, a minimum mean square error (MMSE) reconstruction method is used to retrieve the surface structure under continuous assumption [25]. The object is placed at $\tau$=2.4 in the same scattering media, with one surface facing 20° with the spacing between the two beams. The surface and the reconstruction is shown in figure 14, along with a comparison of normal imaging results (of a black disk object) near τ=1 to indicate the visibility limit.

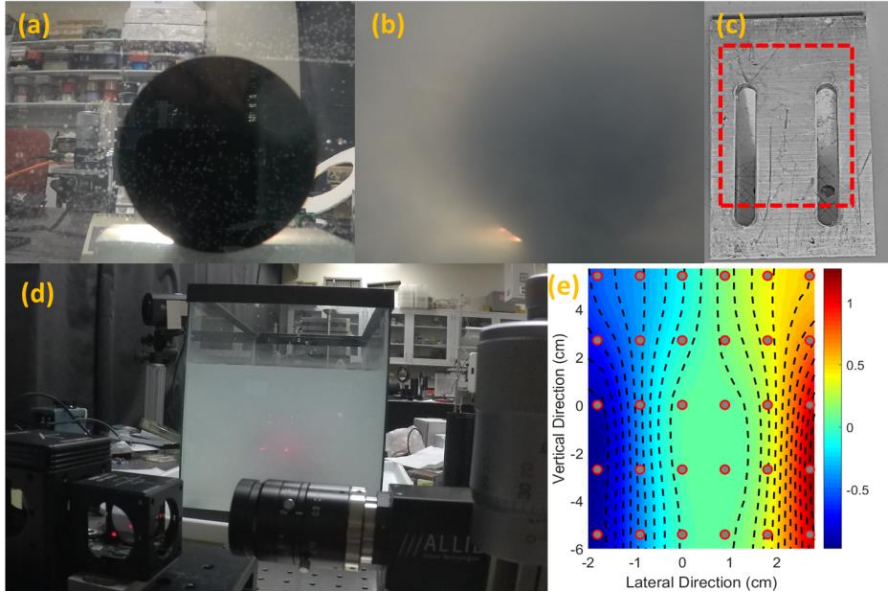

Figure 14: Experimental demonstration of reconstructing a complex tilting surface with sparse scanning pattern: (a) Image of an object in clear water at τ=1.08; (b) Image of the same object in dirty water at τ=1.08 to show the visibility limit; (c) Image of the target and area of interest; (d) Experimental image of laser projection; (e) Reconstruction result.

Figure 14 shows that the asymmetry method can be used to determine the basic surface structure of an object beyond the visibility limit. In plot 14(e), the colormap shows the tilting direction, and the dashed contour lines show the local discrepancies caused by hollow structures in the surface. The red dots in 14(e) represent the projected trajectory center of the twin beams. Evidently, both the general surface tilt and the existence of local structures that deviates from the general tilt can be revealed by the asymmetry method for a hidden object. This preliminary result indicates that the method can be potentially used to profile a hidden object in a scattering environment.

## 4. CONCLUSION AND DISCUSSION

It has been shown through theory and experiments that the backscatter asymmetry from a twin beam projection pattern can be used to extract useful information of a hidden object in a relatively strong light scattering environment. This approach shows that while scattering is often seen as a complication, there are ways to use the inherent optical properties of the channel to detect information about a target such as tilt angles and surface structure variation. Especially when the target becomes invisible due to a significant amount of volumetric back scattered light, the asymmetry method is able to use the differential part of the back scattered signal for object detection and profiling. The method is also compatible with both CW and pulsed lasers, which can either leads to a cost-effective platform (by using CW lasers) or be combined with time-of-flight (TOF) concepts (by using pulse lasers) for enhanced results. When compared to the conventional vision limit of highly scattering underwater environments, such a method can extend the detection range of these features by a factor of 3~5.

It is of great interest to point out the resolution limit of such a system, yet its validation awaits additional experimental data under better alignment conditions. The resolution limit refers to the smallest diameter of a sub-aperture area that is detectable through the asymmetry method. Equivalently, this is the same as the smallest beam spacing $d$ under a given optical thickness τ. The minimum spacing can be expressed as:

$$d_{\min} = \frac{1}{4}\pi w_0 \frac{\exp(\tau)}{\sqrt{\tau}}$$

For the twin beams with beam width of 1mm and τ=1, 2, 3, 4, 5, 6, respectively, the resolution limits are 2.1 mm, 4.1 mm, 9.1 mm, 21.4 mm, 52.1mm and 129 mm. Given that the resolution degrades significantly after τ>6 (assumed that the SAA is still valid in this regime [22]), another methods must be explored to detect objects in turbid underwater environments.